\def\year{2024}\relax
\documentclass[letterpaper]{article} 
\usepackage{aaai24}  
\usepackage{times}  
\usepackage{helvet}  
\usepackage{courier}  
\usepackage[hyphens]{url}  
\usepackage{graphicx} 
\usepackage{natbib}  
\usepackage{caption} 
\usepackage{algorithm}
\usepackage{algorithmic}
\usepackage{mathtools}
\usepackage{booktabs}
\usepackage{bm}
\usepackage{makecell}

\usepackage{newfloat}
\usepackage{listings}
\DeclareCaptionStyle{ruled}{labelfont=normalfont,labelsep=colon,strut=off} 
\floatstyle{ruled}
\newfloat{listing}{tb}{lst}{}
\floatname{listing}{Listing}
\title{AT4CTR: Auxiliary Match Tasks for Enhancing Click-Through Rate Prediction}
\author {
    Qi Liu\textsuperscript{\rm 1} \footnote{This work was done when the author Qi Liu was at Meituan for intern.},
    Xuyang Hou\textsuperscript{\rm 2}, 
    Defu Lian\textsuperscript{\rm 1}\footnote{Corresponding author}, 
    Zhe Wang\textsuperscript{\rm 2}, 
    Haoran Jin\textsuperscript{\rm 1}, 
    Jia Cheng\textsuperscript{\rm 2}
    Jun Lei\textsuperscript{\rm 2}
}
\affiliations {
    \textsuperscript{\rm 1} University of Science and Technology of China\\
    \textsuperscript{\rm 2} Meituan
    
    \{qiliu67,HaoranJin\}@mail.ustc.edu.cn, 
    \{liandefu\}@ustc.edu.cn\\
    \{houxuyang,wangzhe65,jia.cheng.sh,leijun\}@meituan.com\\
}

\usepackage{bibentry}

\begin{document}

\maketitle

\begin{abstract}
Click-through rate (CTR) prediction is a vital task in industrial recommendation systems. Most existing methods focus on the network architecture design of the CTR model for better accuracy and suffer from the data sparsity problem. Especially in industrial recommendation systems, the widely applied negative sample down-sampling technique due to resource limitation worsens the problem, resulting in a decline in performance. In this paper, we propose \textbf{A}uxiliary Match \textbf{T}asks for enhancing \textbf{C}lick-\textbf{T}hrough \textbf{R}ate prediction accuracy (AT4CTR) by alleviating the data sparsity problem. Specifically, we design two match tasks inspired by collaborative filtering to enhance the relevance modeling between user and item. As the "click" action is a strong signal which indicates the user's preference towards the item directly, we make the first match task aim at pulling closer the representation between the user and the item regarding the positive samples. Since the user's past click behaviors can also be treated as the user him/herself, we apply the next item prediction as the second match task. For both the match tasks, we choose the InfoNCE as their loss function. The two match tasks can provide meaningful training signals to speed up the model's convergence and alleviate the data sparsity. We conduct extensive experiments on one public dataset and one large-scale industrial recommendation dataset. The result demonstrates the effectiveness of the proposed auxiliary match tasks. AT4CTR has been deployed in the real industrial advertising system and has gained remarkable revenue.
\end{abstract}

\section{Introduction}
Click-through rate (CTR) prediction is crucial in industrial web applications, e.g. recommendation systems and online advertising. It estimates the probabilities of the user clicking on items and displays the top-ranked items to the user. In online advertising, the platform can only charge the advertiser once the ad is clicked by the user. Thus, accurate CTR estimation can maintain the user's satisfaction and maximize the revenue for both the platform and the advertiser. 

Existing advance in CTR mainly focuses on network architecture and have gained huge success. Traditional methods, like Logistic Regression~\cite{richardson2007predicting} and Factorization Machine (FM)~\cite{rendle2010factorization}, can only capture the low-order feature interactions. Recently, deep learning has been exploited for CTR prediction. Methods, such as Wide\&Deep~\cite{cheng2016wide}, DeepFM~\cite{guo2017deepfm}, and DCN~\cite{wang2017deep}, focus on capturing the high-order feature interactions through the neural network. On the other side, DIN~\cite{zhou2018deep}, DIEN~\cite{zhou2019deep}, and DBPMaN~\cite{dong2023deep} extract user interest from user behavior sequences like click or conversion. Those methods all improve the performance of CTR prediction by a large margin.

The success of network architecture makes researchers ignore another important problem of data sparsity which means that positive samples take up only a small part of the total samples. Especially in the industrial scenarios, negative sample down-sampling which abandons each negative sample based on the Bernoulli distribution with certain probability is widely used to reduce the computing and storage cost when training the CTR model. However, drastically abandoning negative samples will worsen the data sparsity issue and degrade the performance. As there always exists abandoned hard negative samples which are important for the CTR model's updating. A few works devote efforts to solving the data sparsity issue. DeepMCP~\cite{ouyang2019representation} applies a matching subnet to strengthen the relevance between the user and the item, and a correlation subnet to improve item representation. But it introduces no extra training signals. DMR~\cite{lyu2020deep} designs an auxiliary network to predict the last behavior based on previous behaviors. The auxiliary losses used in DeepMCP and DMR are all negative sampling~\cite{mikolov2013distributed} which is an approximation of full softmax. It is not strong enough and implementation unfriendly as the item has a large magnitude and changes over time. CL4CTR~\cite{wang2023cl4ctr} exploits an auxiliary network to perform self-supervised contrastive learning but triples the training cost. 
 
To alleviate the data sparsity problem, we propose two novel auxiliary match tasks to provide more helpful training signals. As shown in Figure~\ref{fig:at4ctr_framework}, it contains two auxiliary tasks. The "click" action demonstrates the user's strong preference for the clicked item. Intuitively, the representation of positive samples between the user side and the item side features should be highly relevant. The main binary classification task can not fully concentrate on user-item relevance learning because it needs to deal with the context and other interaction features simultaneously~\cite{lin2023map}. Thus, we propose the first auxiliary match task named User-Item Match (UIM). Inspired by the success of contrastive learning in CV\&NLP, we take the InfoNCE~\cite{oord2018representation} as the loss of the UIM task. It treats each positive sample's user and item as positive pairs and pulls closer their representation. It takes the other samples in the same batch to synthesize negative samples whose representation of user and item should be pushed away. The UIM task provides explicit signals to model the relevance between the user and the item, which relieves the stress of the main task and speeds up convergence. The behavior sequence consists of the user's clicked items in chronological order, which contains the causality of the user's decision and can represent himself/herself to some extent. We design the second auxiliary match task named Next Item Prediction (NIP). Specifically, we aggregate the previous behaviors through self-attention~\cite{vaswani2017attention} as the user's representation and predict the next item, which can also be regarded as a micro UIM task. We exploit the InfoNCE as an approximation of full softmax due to the large magnitude of items. The NIP task can accelerate the convergence of the behavior sequence modeling module. We choose InfoNCE because of its inherent ability to mine hard negative samples\cite{wang2021understanding}. This ability can release the power of the synthetic negative samples from the in-batch negative construction and benefits the negative sample down-sampling industrial scenarios more by supplementing more hard negative signals. In summary, the main contributions of this paper are:
\begin{itemize}
    \item We reveal the neglected problem of data sparsity which is serious in the negative sample down-sampling industrial scenarios and propose two auxiliary match tasks to provide extra meaningful training signals.
    \item We propose the AT4CTR, which contains two auxiliary match tasks to strengthen the relevance between the user and the item with the help of InfoNCE loss.
    \item We conduct offline experiments on both public and industry datasets to verify the effectiveness of the proposed AT4CTR, which achieves remarkable improvement in the online A/B test.
\end{itemize}

\section{Related Work}
\subsection{Deep CTR Modeling}
Recent CTR research based on deep learning can be mainly divided into two directions: feature interaction and user behavior sequence modeling. The feature interaction methods believe that the interaction between different features is important for CTR modeling. Early FM-based methods only model the second-order pairwise interactions by using factorized parameters, which limits the performance of CTR modeling. Many works explore how to capture high-order and informative feature interactions efficiently. Wide\&Deep (WDL)~\cite{cheng2016wide} exploits Deep Neural Network (DNN) to capture the high-order feature interaction implicitly for capturing high-order feature interaction. DeepFM~\cite{guo2017deepfm} combines DNN and FM, and xDeepFM~\cite{lian2018xdeepfm} further proposes the Compressed Interaction Network to model the high-order feature interaction explicitly. DCN~\cite{wang2017deep} and DCN-V2~\cite{wang2021dcn} apply cross-vector/matrix network to achieve informative feature interaction automatically. 

User behavior sequence modeling is another important part of CTR modeling. It focuses on extracting the user's interest from the behavior sequence which is composed of interacted items by the user in chronological order. DIN~\cite{zhou2018deep} first applies the attention mechanism to mine user interest by activating items related to the target item and gains huge performance improvement. Based on DIN, DIEN~\cite{zhou2019deep} utilizes a two-layer GRU to capture the dynamic change of the user's interest. Works~\cite{pi2020search,chang2023twin,lin2022sparse} propose to extract long-term interest from the user's ultra-long behavior sequence by taking the approximate nearest neighbor search algorithm to reduce latency. Some works~\cite{guo2019buying,zhou2018atrank} introduce multiple types of behavior sequences to obtain fine-grained user interest. DBPMaN~\cite{dong2023deep} proposes a new perspective for behavior sequence modeling. It introduces the concept of behavior path to understand the psychological procedure behind the user's decision. However, all the above methods devote too much effort to the network structure's design and ignore the data sparsity problem.

\subsection{Contrastive Learning for Recommendation}
Contrastive learning is a self-supervised learning algorithm, aiming to obtain invariant representation by optimizing the goal of mutual information maximization, and gains huge success in CV\&NLP~\cite{gao2021simcse,chen2020simple}. Recently, some works have introduced contrastive learning into the recommendation system. In sequential recommendation, task~\cite{zhou2021contrastive,xie2022contrastive,zhang2023fairlisa}, the augmented user's behavior sequence, produced by inserting, masking, shuffling, etc, is treated as a positive pair. The additional contrastive learning task enhances the representation learning ability of the recommendation model and thus gains performance improvement. In the CTR prediction task, contrastive learning has not been well explored. MISS~\cite{guo2022miss} focuses on sequential-based CTR tasks, which apply interest-level contrastive learning to enhance the behavior sequence modeling. CL4CTR~\cite{wang2023cl4ctr} improves the quality of feature representation by designing three self-supervised tasks: contrastive learning, feature alignment constraint, and field uniformity constraint. However, the proposed three self-supervised tasks are unrealistic for industrial CTR model training because they need to regularize the huge embedding table and triple the training overhead at least. AT4CTR also exploits contrastive learning but can enhance performance efficiently with little extra training cost. 

\section{Auxiliary Match Tasks for CTR}
\begin{figure*}
    \centering
    \includegraphics[width=1.0\textwidth,height=0.5\textwidth]{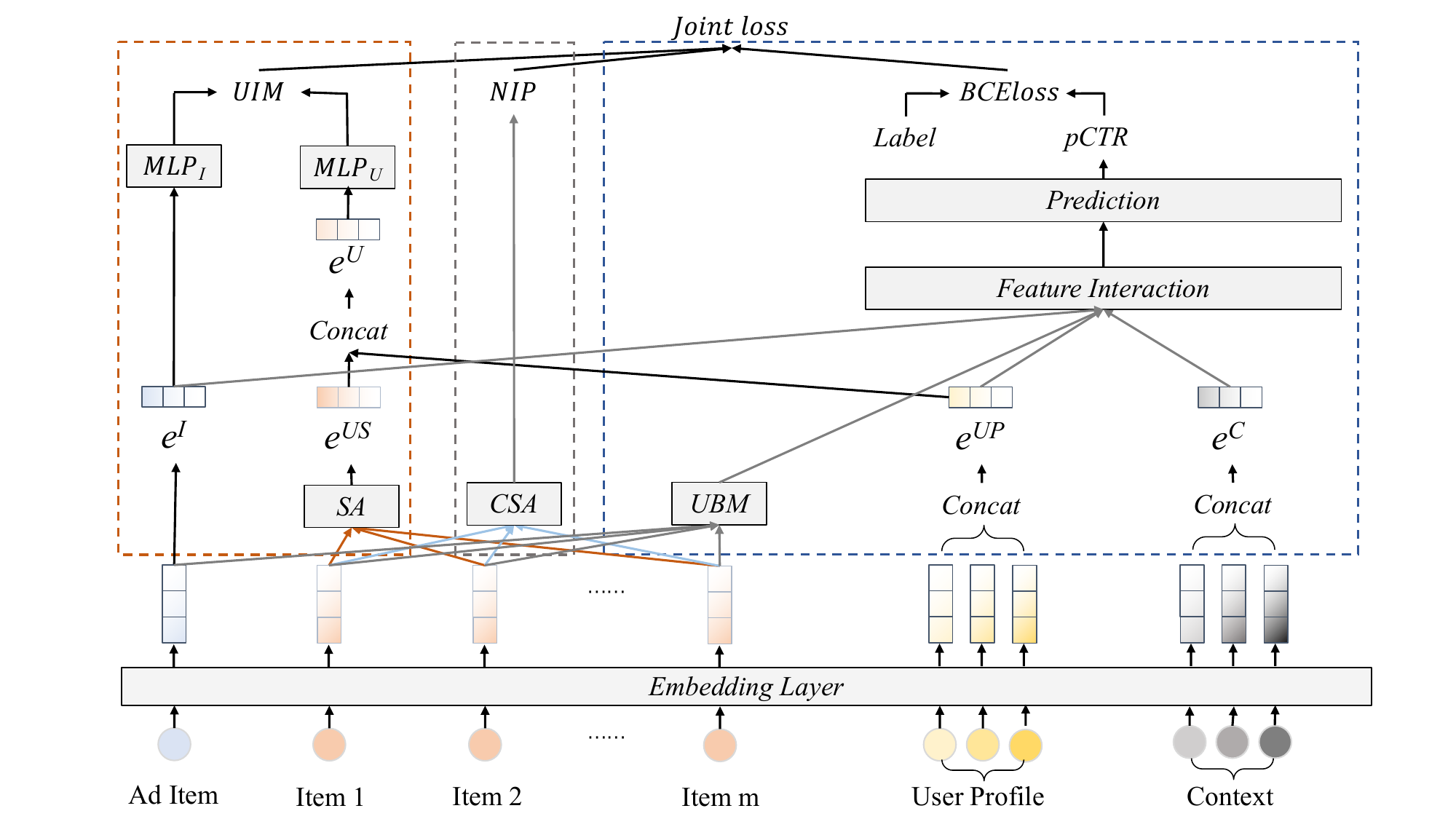}
    \caption{The overall framework of AT4CTR. \textit{SA} represents self-attention and \textit{CSA} indicates causal self-attention. \textit{UBM} means user behavior sequence modeling. \textit{UIM} and \textit{NIP} are the two proposed auxiliary match tasks.}
    \label{fig:at4ctr_framework}
\end{figure*}

\subsection{Overview}
CTR modeling is a binary classification machine learning task based on sparse features. The sparse features mainly contain information about the user, item, and context. The user side information contains the user's profile (e.g., age, gender, city) and the user's behavior sequence. The item side information consists of item\_id, category\_id, etc. The context information is composed of position, time, etc. The CTR model needs to estimate the probability of the user clicking on the item under the given context. We represent the instance by $\{\mathbf{x}, y\}$, where $\mathbf{x}=[\mathbf{x^{UP}}, \mathbf{x^{US}}, \mathbf{x^I}, \mathbf{x^C}]$, $y\in\{0,1\}$ indicates click or not. $UP, US, I$, and $C$ represent the features' set of user profile, user behavior sequence, item, and context respectively. The CTR task can be formulated as the following Eq~(\ref{eq:ctr_formulation}):
\begin{equation}
    \label{eq:ctr_formulation}
    P(y=1|\mathbf{x})=F(\mathbf{x})
\end{equation}
where $F$ is the CTR model.

AT4CTR takes $\mathbf{x}$ as input and transforms it into dense vector through the embedding layer.
As shown in Figure~\ref{fig:at4ctr_framework}, the UIM task takes the $\mathbf{x^{UP}}, \mathbf{x^{US}}$ and $\mathbf{x^{I}}$ as input. It applies an independent self-attention~\cite{vaswani2017attention} network to aggregate behavior sequence. The user representation is composed of the aggregated behavior sequence and user profile. Then the InfoNCE loss strengthens the relevance between the user and the item of positive samples. NIP task exploits another self-attention network to perform the causal behavior sequence aggregation based on $\mathbf{x^{US}}$. And the Infonce loss is applied to supervise the process. The two auxiliary match tasks only add little training cost and do not increase the parameters and latency during inference. 

\subsection{Embedding Layer}
The embedding layer transforms the high-dimensional sparse vector $\mathbf{x}$ into low-dimensional dense representations. Specifically, each feature field will be assigned with an embedding matrix $\mathbf{E=[e_1;e_2;...;e_K]} \in R^{K \times d}$, where $K$ represents the cardinality of this feature field and $d$ donates the embedding size. If the index value of the feature is $i$, then $\mathbf{e_i}$ serves as its embedding. 

\subsection{UIM: User-Item Match Task}
The UIM which applies InfoNCE loss aims at pulling closer the representation between user and item of the positive sample. To avoid representation collapse, we combine the user/item in the positive samples with the item/user of the other samples in the same batch as the negative term of the InfoNCE loss. The representation between the user and the item of synthetic negative samples will be pushed away.

Specifically, the representation of the user side contains the embedding of user profile and the aggregated behavior sequence. There are multiple types of features in $\mathbf{x^{UP}}$, we concatenate the embedding of all these features and get the user profile representation $\mathbf{e^{UP}}$. For the behavior sequence, we first apply self-attention~\cite{vaswani2017attention} to refine behavior sequence representation as it can capture the relatedness between the clicked items as Eq~(\ref{eq:sa}):
\begin{equation}
    \label{eq:sa}
    SA(X)=Softmax(\frac{XW^Q({XW^K})^T}{\sqrt{d}})XW^V
\end{equation}
where $X$ is the user's behavior sequence $\mathbf{x^{US}}$, $W^Q, W^K, W^V \in R^{d \times d}$ are the weight matrix to generate query, key, and value respectively.
After that, we perform the mean pool aggregation operation to get the representations of the user's interest $\mathbf{e^{US}}$ as Eq~(\ref{eq:mean_pooling}):
\begin{equation}
    \label{eq:mean_pooling}
    \mathbf{e^{US}}=mean\_pool(SA(\mathbf{x^{US}}))
\end{equation}
We then concat the user profile representation and the user's interest as the user's representation $\mathbf{e^U}$. 
\begin{equation}
    \label{eq:eu}
    \mathbf{e^{U}}=concat(\mathbf{e^{UP},e^{US}})
\end{equation}
For the representation of the item side, we concatenate the embeddings of features in $\mathbf{x^I}$ to get the item representation $\mathbf{e^I}$. The number of features differs between the user side and the item side. We apply two separate Multi-Layer Perceptron (a.k.a Projection Head) ~\cite{chen2020simple} to align their representation, which follows the Eq(~\ref{eq:up_proj}):
\begin{equation}
    \label{eq:up_proj}
    \begin{split}
        &\mathbf{r^U}=MLP_u(\mathbf{e^U}),\\
        &\mathbf{r^I}=MLP_I(\mathbf{e^I})
    \end{split}
\end{equation}
where $MLP_u$ and $MLP_I$ are all two-layer MLP with ReLU as activation, $\mathbf{r^U}$ and $\mathbf{r^I}$ are the aligned representations. Then we apply the InfoNCE for all positive samples based on the aligned representations as Eq(~\ref{eq:up_infonce}).
\begin{equation}
    \label{eq:up_infonce}
    L_{ui}=-\frac{1}{n_+}\sum_{k=1}^{n_+}\log\frac{\exp(sim(\mathbf{r_+^U, r_+^I})/\tau_1)}{\sum_{j=1}^n\exp(sim(\mathbf{r_+^U, r_j^I})/\tau_1)}
\end{equation}
where $sim(\cdot)$ represents the cosine similarity, $\tau_1$ is the temperature hyperparameter, $n$ is the batch size, and $n_+$ is the number of positive samples in the batch. The InfoNCE loss should be symmetrical, so we also compute the $L_{iu}$ as follows Eq(~\ref{eq:pu_infonce}):
\begin{equation}
    \label{eq:pu_infonce}
    L_{iu}=-\frac{1}{n_+}\sum_{k=1}^{n_+}\log\frac{\exp(sim(\mathbf{r_+^I, r_+^U})/\tau_1)}{\sum_{j=1}^n\exp(sim(\mathbf{r_+^I, r_j^U})/\tau_1)}
\end{equation}
Combining the two losses, we obtain the UIM auxiliary loss as $L_{UIM}=L_{ui}+L_{iu}$.

\subsection{NIP: Next Item Prediction Task}
The user behavior sequence which consists of interacted items contains the causal psychological clues of the user implicitly. One obvious intuition is that the past behaviors of the user will affect the current behavior. We design the second auxiliary match task which performs the next item prediction task by taking InfoNCE as an approximation of full softmax. From another side, we can also treat the past behaviors as the user's representation and the next behavior as the positive target item, which is consistent with the mechanism of the UIM.

Specifically, we take the auto-regressive causal self-attention~\cite{vaswani2017attention} as the encoder of behavior sequence in NIP as Eq~(\ref{eq:csa}):
\begin{equation}
    \label{eq:csa}
    [\mathbf{r^1,r^2,..,r^{m-1}}] = Causal\_SA(\mathbf{x^{US}})
\end{equation}
where the $\mathbf{r^t}, t\in{\{1,2,.., m-1\}}$ is the aggregated representation of oldest $t$ behaviors. We treat the embedding $\mathbf{e^I}$ of the next item as the representation $\mathbf{r^I}$ of the next item. To calculate the InfoNCE loss, we still apply the in-batch negative construction. Items at the same position in the behavior sequence in other samples of the same batch are regarded as negative samples. 

We calculate InfoNCE loss of the next item prediction task as Eq~(\ref{eq:pi_infonce}):
\begin{equation}
    \label{eq:pi_infonce}
    L_{pi}=-\frac{1}{n(m-1)}\sum_{i=1}^{n}\sum_{k=1}^{m-1}\log\frac{\exp(sim(\mathbf{r_i^k, e_i^{k+1}})/\tau_2)}{\sum_{j=1}^n\exp(sim(\mathbf{r_i^k, e_j^{k+1}})/\tau_2)}
\end{equation}
where $n$ is the batch size, m is the length of behavior sequence, $\tau_2$ is the temperature hyperparameter, $\mathbf{r_i^k}$ is the aggregated representation of the first $k$ behaviors, and $\mathbf{e_i^{k+1}}$ is the embedding of the $k+1$-th item. We also have the symmetrical InfoNCE loss $L_{ip}$ as Eq(~\ref{eq:ip_infonce}).
\begin{equation}
    \label{eq:ip_infonce}
    L_{ip}=-\frac{1}{n(m-1)}\sum_{i=1}^{n}\sum_{k=1}^{m-1}\log\frac{\exp(sim(\mathbf{e_i^{k+1}, r_i^k})/\tau_2)}{\sum_{j=1}^n\exp(sim(\mathbf{e_i^{k+1}, r_j^k})/\tau_2)}
\end{equation}
Adding both losses together, we get the loss of the auxiliary NIP task as $L_{NIP}=L_{pi}+L_{ip}$.

\subsection{Multi-task Training}
We use the widely applied negative log-likelihood loss as the main loss of CTR prediction as Eq~(\ref{eq:bce_loss}):
\begin{equation}
    \label{eq:bce_loss}
    L_{main}=-y\log F(x)-(1-y)(1-\log F(x))
\end{equation}
We add the main loss and losses of the two auxiliary tasks together to supervise the model training. The total loss is as Eq~(\ref{eq:loss_total}):
\begin{equation}
    \label{eq:loss_total}
    L_{total}=L_{main}+\lambda_{UIM}L_{UIM}+\lambda_{NIP}L_{NIP}
\end{equation}
where the $\lambda_{UIM}$ and $\lambda_{NIP}$ are the weight coefficients.

\section{Experiment Settings}
\subsection{Datasets}
We conduct extensive experiments on both the industry and the public datasets. The statistics information about the two datasets are shown in Table~\ref{tab:data_stats}.
\subsubsection{Taobao Dataset}
The Taobao dataset~\cite{zhu2018learning} is widely used in CTR research. It consists of a set of user behaviors from Taobao's industry recommendation system. The dataset contains about 1 million users whose behaviors include clicking, purchasing, adding items to the shopping cart, etc. The click behaviors for each user are taken and sorted according to the timestamp to construct the behavior sequence. We filter out users who have less than 10 behaviors. The split standard is the same as what CAN~\cite{bian2022can} does. 

\subsubsection{Industry Dataset}
We collect traffic logs from the search advertising system in the location-based service e-commerce platform of Meituan. The last ten months' samples are used for training and samples of the following day are for testing. For the training set, we perform negative sample down-sampling with the ratio $0.1$. As the testing set is only for evaluating the offline metric, we don't perform negative sample down-sampling anymore. Following~\cite{he2014practical}, we re-calibrate the model for the online severing.

\begin{table}[tb!]
\renewcommand\arraystretch{0.85}
\caption{Statistics of datasets.}
\centering
\label{tab:data_stats}
\resizebox{1.\columnwidth}{!}{
\begin{tabular}{cl|c|c|c|c}
\toprule
\multicolumn{2}{c|}{Datasets}   & \multicolumn{1}{c}{\#Users} & \multicolumn{1}{c}{\#Items} & \multicolumn{1}{c}{\#Fields} & \multicolumn{1}{c}{\#Instances} \\
\midrule
\multicolumn{2}{c|}{Taobao}           & \multicolumn{1}{c}{987,991}             & \multicolumn{1}{c}{4,161,138}            & \multicolumn{1}{c}{7}               & \multicolumn{1}{c}{100,095,182}  \\
\multicolumn{2}{c|}{Industry}           & \multicolumn{1}{c}{40M}             & \multicolumn{1}{c}{417K}            & \multicolumn{1}{c}{168}               & \multicolumn{1}{c}{6.6B}  \\
\bottomrule 
\end{tabular}
}
\end{table}

\subsection{Baselines}
We compare AT4CTR with three types of CTR modeling methods. The first type are feature interaction methods including \textbf{FM}~\cite{rendle2010factorization}, \textbf{WideDeep}~\cite{cheng2016wide}, \textbf{DeepFM}~\cite{guo2017deepfm}, which focus on second and high order feature interactions. The second type is the behavior sequence modeling method. \textbf{DIN}~\cite{zhou2018deep} uses the attention mechanism to extract the user's candidate-aware interest. \textbf{DIEN}~\cite{zhou2019deep} extends DIN with the interest extractor layer and interest evolution layer. \textbf{DBPMaN}~\cite{dong2023deep} exploits the behavior path to capture the psychological procedure behind user decisions. The third type is the auxiliary task method. \textbf{DeepMCP}~\cite{ouyang2019representation} uses the matching subnet to capture the user-item relation, and the correlation subnet to explore the item-item correlation. \textbf{DMR}~\cite{lyu2020deep} proposes an auxiliary matching loss to measure the correspondence between the user preference and the target item in the behavior sequence. \textbf{CL4CTR}~\cite{wang2023cl4ctr} exploits an auxiliary network to perform contrastive learning, which aims to enhance the embedding representation.


\subsection{Evaluation Metric}
Two widely used metrics AUC, and Logloss are chosen. The AUC (Area Under the ROC Curve) measures the comprehensive ranking ability of the CTR model for all samples in the testing set. The Logloss measures the accuracy of the estimated probability depending on the ground-truth label. A slight improvement of AUC or Logloss at $\mathbf{0.001}$-level is significant in a mature recommendation system~\cite{guo2017deepfm}, as it means a huge promotion in revenue. 

\subsection{Implementation Details}
We implement AT4CTR with Tensorflow. For the industrial dataset, the embedding size is $16$ and the learning rate is $5e-4$. We train the model using $8$ $80G$ $A100$ GPUs with the batch size 1500 of a single card. For the Taobao dataset, we set the embedding size to be $18$, the learning rate to be $1e-3$, and use one single $80$ $A100$ for training with batch size $1024$. We set $\tau_1$ to be $0.07$ and $\tau_2$ to be $0.1$. We use Adam as the optimizer for both datasets. We run all experiments five times and report the average result. For DeepMCP and CL4CTR, we exploit the DBPMaN to do behavior sequence modeling.

\section{Experiment Results}
\subsection{Performance Comparison}
\begin{table}[tb!]
\centering
\caption{Performance comparison of baselines on two datasets. The best result is in boldface and the second best is underlined. * indicates that the difference to the best baseline is statistically significant at 0.01 level.}
\label{tab:main_result}
\begin{tabular}{cl|c|c|c|c}
\toprule
& & \multicolumn{2}{c}{Industry} &\multicolumn{2}{c}{Taobao} \\
& & \multicolumn{1}{c}{AUC} &\multicolumn{1}{c}{Logloss} &\multicolumn{1}{c}{AUC} &\multicolumn{1}{c}{Logloss} \\
\midrule
\multicolumn{2}{c|}{FM}        & \multicolumn{1}{c}{0.7177}   & \multicolumn{1}{c|}{0.1955}   & \multicolumn{1}{c}{0.8025}   & \multicolumn{1}{c}{0.2723} \\
\multicolumn{2}{c|}{WideDeep}  & \multicolumn{1}{c}{0.7335}   & \multicolumn{1}{c|}{0.1923}   & \multicolumn{1}{c}{0.8733}   & \multicolumn{1}{c}{0.2233} \\
\multicolumn{2}{c|}{DeepFM}    & \multicolumn{1}{c}{0.7327}   & \multicolumn{1}{c|}{0.1924}   & \multicolumn{1}{c}{0.8690}   & \multicolumn{1}{c}{0.2263} \\
\multicolumn{2}{c|}{DIN}       & \multicolumn{1}{c}{0.7420}   & \multicolumn{1}{c|}{0.1906}   & \multicolumn{1}{c}{0.9402}   & \multicolumn{1}{c}{0.1544} \\
\multicolumn{2}{c|}{DIEN}      & \multicolumn{1}{c}{0.7424}   & \multicolumn{1}{c|}{0.1905}   & \multicolumn{1}{c}{0.9479}   & \multicolumn{1}{c}{0.1430} \\
\multicolumn{2}{c|}{DBPMaN}    & \multicolumn{1}{c}{0.7426}   & \multicolumn{1}{c|}{0.1905}   & \multicolumn{1}{c}{0.9509}   & \multicolumn{1}{c}{0.1382} \\
\multicolumn{2}{c|}{DeepMCP}   & \multicolumn{1}{c}{0.7426}   & \multicolumn{1}{c|}{0.1906}   & \multicolumn{1}{c}{\underline{0.9514}}   & \multicolumn{1}{c}{\underline{0.1376}} \\
\multicolumn{2}{c|}{DMR}       & \multicolumn{1}{c}{0.7421}   & \multicolumn{1}{c|}{0.1906}   & \multicolumn{1}{c}{0.9405}   & \multicolumn{1}{c}{0.1540} \\
\multicolumn{2}{c|}{CL4CTR}    & \multicolumn{1}{c}{\underline{0.7428}}   & \multicolumn{1}{c|}{\underline{0.1904}}   & \multicolumn{1}{c}{0.9508}  & \multicolumn{1}{c}{0.1384} \\  \midrule
\multicolumn{2}{c|}{AT4CTR}    & \multicolumn{1}{c}{\bm{$0.7441^*$}}   & \multicolumn{1}{c|}{\bm{$0.1902^*$}}   & \multicolumn{1}{c}{\bm{$0.9535^*$}}   & \multicolumn{1}{c}{\bm{$0.1345^*$}} \\ 
\bottomrule 
\end{tabular}
\end{table}
Table~\ref{tab:main_result} shows the results of all methods. AT4CTR obtains the best performance on both the Industry and Taobao datasets, which shows the effectiveness of AT4CTR. There are some insightful findings from the results. 
(1) The proposed AT4CTR beats all baselines on both metrics on both datasets. Compared with methods of feature interaction and behavior sequence modeling, AT4CTR forces the neural network to capture the relevance between user and item through two proposed auxiliary tasks. This demonstrates that the proposed auxiliary tasks based on the intrinsic character of the data itself can alleviate the data sparsity issue and improve the training of the model. 
(2) FM performs worse than WideDeep and other deep learning models significantly, which reveals the importance of non-linear transformation and high-order feature interactions. 
(3) From Table~\ref{tab:main_result}, behavior sequence modeling methods outperform feature interaction methods significantly, which verifies the necessity of extracting user interest. DIN extracts the user interest by considering the relevance of the behavior sequence with regard to the target item but ignores the sequential character of user behavior. DIEN applies the two-layer GRU structure and the auxiliary binary classification task to capture the evolution of user interest, and thus performs better than DIN. DBPMaN exploits the behavior path to understand the psychological procedure behind user decisions and achieves the best performance among behavior sequence modeling methods. 
(4) Auxiliary task methods also benefit the performance. The DeepMCP uses a matching subnet to strengthen the correlation between user and item through binary classification task and takes the skip-gram algorithm to model items' correlation. However, it can only achieve slight improvement and we think the reason is that its auxiliary binary classification task is homogeneous with the main CTR task, which can not provide extra signals. For DMR, its auxiliary task is to predict the last behavior based on the previous behaviors with the negative sampling technique to approximate the multi-classification task. It provides weak training signal and obtains little earnings. The CL4CTR shows almost no improvement on both datasets.
 
\subsection{Ablation Study}
\begin{table}[tb!]
\centering
\caption{Effect of each auxiliary task.}
\label{tab:aux_task_ablation}
\begin{tabular}{cl|c|c|c|c}
\toprule
& & \multicolumn{2}{c}{Industry} &\multicolumn{2}{c}{Taobao} \\
& & \multicolumn{1}{c}{AUC} &\multicolumn{1}{c}{Logloss} &\multicolumn{1}{c}{AUC} &\multicolumn{1}{c}{Logloss} \\
\midrule
\multicolumn{2}{c|}{DBPMaN}    & \multicolumn{1}{c}{0.7426}   & \multicolumn{1}{c|}{0.1905}   & \multicolumn{1}{c}{0.9509}   & \multicolumn{1}{c}{0.1382} \\
\multicolumn{2}{c|}{+UIM}      & \multicolumn{1}{c}{0.7438}   & \multicolumn{1}{c|}{0.1903}   & \multicolumn{1}{c}{0.9529}   & \multicolumn{1}{c}{0.1360} \\  \midrule
\multicolumn{2}{c|}{\makecell[c]{DBPMaN\\ (+UIM,+NIP)}} & \multicolumn{1}{c}{\bm{$0.7441$}}   & \multicolumn{1}{c|}{\bm{$0.1902$}}   & \multicolumn{1}{c}{\bm{$0.9535$}}   & \multicolumn{1}{c}{\bm{$0.1345$}} \\ 
\bottomrule 
\end{tabular}
\end{table}
We investigate how the UIM and NIP auxiliary tasks influence AT4CTR and show the results in Table~\ref{tab:aux_task_ablation}. The ablation experiments are conducted based on the DBPMaN. We first integrate UIM with DBPMaN. And then based on UIM, we further combine NIP with DBPMaN. From Table~\ref{tab:aux_task_ablation}, both the auxiliary tasks are beneficial for the DBPMaN on both datasets, which indicates their ability to alleviate data sparsity problem. The UIM works by directly capturing the fine-grained relevance between the user and the item, which is not easy for the main task as it needs to provide compromised gradients for all features. For the NIP, it models the items' correlation of behavior sequence. However, the information of each behavior which only contains item\_id, category\_id, etc in the behavior sequence is limited due to the storage cost. This explains why the benefit enhanced by NIP is less than UIM, the latter contains plenty of features.

\subsection{The Generalization Ability of AT4CTR}
\begin{table}[tb!]
\caption{Results of combining AT4CTR with different CTR Models, $_{AT}$ means combination.}
\label{tab:combine_models}
\begin{tabular}{cl|c|c|c|c}
\toprule
& & \multicolumn{2}{c}{Industry} &\multicolumn{2}{c}{Taobao} \\
& & \multicolumn{1}{c}{AUC} &\multicolumn{1}{c}{Logloss} &\multicolumn{1}{c}{AUC} &\multicolumn{1}{c}{Logloss} \\
\midrule
\multicolumn{2}{c|}{WideDeep}           & \multicolumn{1}{c}{0.7335}            & \multicolumn{1}{c|}{0.1923}           & \multicolumn{1}{c}{0.8733}            & \multicolumn{1}{c}{0.2233} \\
\multicolumn{2}{c|}{WideDeep$_{AT}$}  & \multicolumn{1}{c}{\bm{$0.7389$}}     & \multicolumn{1}{c|}{\bm{$0.1912$}}    & \multicolumn{1}{c}{\bm{$0.8781$}}     & \multicolumn{1}{c}{\bm{$0.2189$}} \\
\midrule
\multicolumn{2}{c|}{DeepFM}             & \multicolumn{1}{c}{0.7327}                  & \multicolumn{1}{c|}{0.1924}                 & \multicolumn{1}{c}{0.8690}                  & \multicolumn{1}{c}{0.2263} \\
\multicolumn{2}{c|}{DeepFM$_{AT}$}    & \multicolumn{1}{c}{\bm{$0.7375$}}           & \multicolumn{1}{c|}{\bm{$0.1916$}}          & \multicolumn{1}{c}{\bm{$0.8802$}}           & \multicolumn{1}{c}{\bm{$0.2174$}} \\
\midrule
\multicolumn{2}{c|}{DIN}                & \multicolumn{1}{c}{0.7420}                  & \multicolumn{1}{c|}{0.1906}                 & \multicolumn{1}{c}{0.9402}                  & \multicolumn{1}{c}{0.1544} \\
\multicolumn{2}{c|}{DIN$_{AT}$}       & \multicolumn{1}{c}{\bm{$0.7433$}}           & \multicolumn{1}{c|}{\bm{$0.1904$}}          & \multicolumn{1}{c}{\bm{$0.9430$}}           & \multicolumn{1}{c}{\bm{$0.1509$}} \\
\midrule
\multicolumn{2}{c|}{DIEN}               & \multicolumn{1}{c}{0.7423}                  & \multicolumn{1}{c|}{0.1905}                 & \multicolumn{1}{c}{0.9479}                  & \multicolumn{1}{c}{0.1430} \\
\multicolumn{2}{c|}{DIEN$_{AT}$}      & \multicolumn{1}{c}{\bm{$0.7434$}}           & \multicolumn{1}{c|}{\bm{$0.1903$}}          & \multicolumn{1}{c}{\bm{$0.9506$}}           & \multicolumn{1}{c}{\bm{$0.1394$}} \\
\midrule
\multicolumn{2}{c|}{DBPMaN}                & \multicolumn{1}{c}{0.7426}                  & \multicolumn{1}{c|}{0.1905}                 & \multicolumn{1}{c}{0.9509}                  & \multicolumn{1}{c}{0.1382} \\
\multicolumn{2}{c|}{DBPMaN$_{AT}$}       & \multicolumn{1}{c}{\bm{$0.7441$}}           & \multicolumn{1}{c|}{\bm{$0.1902$}}          & \multicolumn{1}{c}{\bm{$0.9536$}}           & \multicolumn{1}{c}{\bm{$0.1344$}} \\
\bottomrule 
\end{tabular}
\end{table}
Since we focus on designing auxiliary tasks for CTR prediction which is orthogonal to most existing CTR prediction models. We show the result by combining AT4CTR with different CTR prediction models. For methods of feature interaction WideDeep and DeepFM, we only combine UIM with them. For the rest behavior sequence modeling methods, we integrate both the UIM and the NIP with them. The result in Table~\ref{tab:combine_models} shows that AT4CTR can boost various CTR models' performance and demonstrates its generalization ability. We have two findings from the result. First, the feature interaction methods gain huge improvements when combined with AT4CTR. The absence of behavior sequence makes the features of user profile gain unexpected stress on user representation. The original binary click/non-click signals together with the feature interaction components (e.g., FM, DNN) are not enough to provide sufficient signals for the representation and relevance learning. The UIM gives an explicit signal to pull closer the representation between user and item with regard to the positive samples and push away the representation of user and item for the synthetic negative samples in InfoNCE loss. The result demonstrates the UIM improves the embedding representation of the user and item features. Then the improved embedding representation eases the learning of feature interaction components (e.g., FM, DNN). Second, when combining both the auxiliary tasks with methods of behavior sequence modeling, performance is also enhanced. Since the mechanism of extracting user interest through the attention network between the target item and the behavior sequence is similar to the effect of two auxiliary tasks. The effect is to constrain the representation between user and item according to given signals, the improvement of AT4CTR here is less than that of the feature interaction methods. All the results indicate that only the main binary classification task can't fully unleash the potential of the CTR model. The proposed auxiliary tasks always help the model's training.

\subsection{The Influence of Negative Down-sampling Ratio}
\begin{figure}
    \centering
    \includegraphics[width=1.0\linewidth]{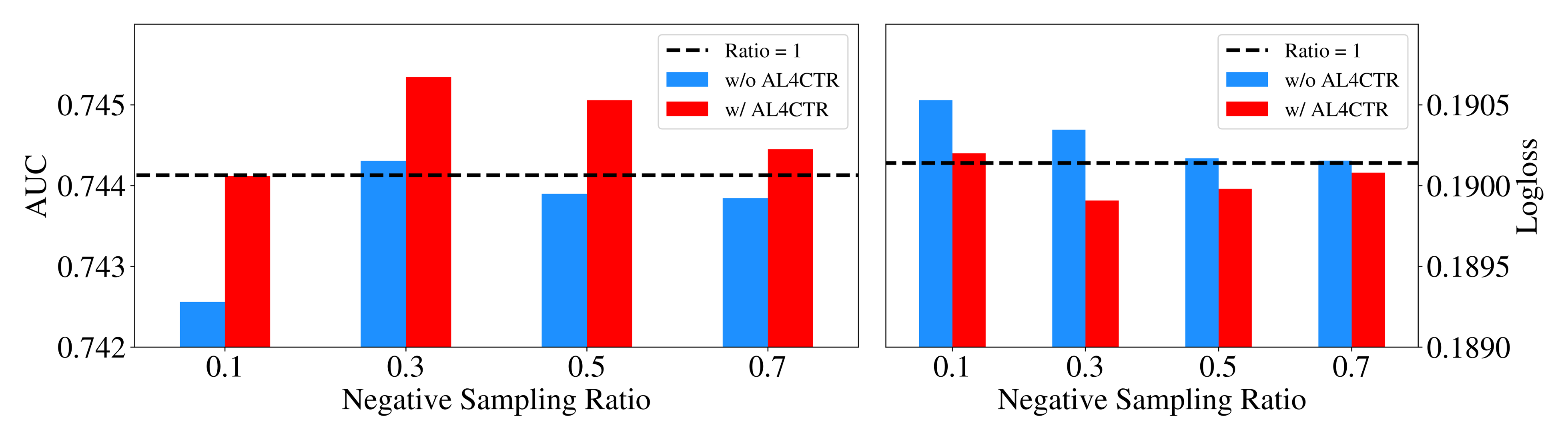}
    \caption{Result of different negative sample down-sampling ratios on Industry dataset.}
    \label{fig:ablation_sampling_ratio}
\end{figure}
As we have claimed the applied InfoNCE loss can provide plenty of synthetic negative samples to make up for the losing performance caused by negative sample downsampling. In this subsection, we perform ablation studies to observe the influence of different negative sampling ratios on the AT4CTR. We present the result of different negative sampling ratios in Figure~\ref{fig:ablation_sampling_ratio}. From Figure~\ref{fig:ablation_sampling_ratio}, we have the following findings. (1) The proposed AT4CTR can enhance the performance under all negative sampling ratios. (2) Under the negative sampling ratio $0.1$, the AT4CTR achieves the same performance as the situation without negative sampling. What's more, the relative promotion enhanced by AT4CTR decreases as the negative sampling ratio increases. This phenomenon is consistent with our hypothesis. As the negative sampling ratio increases, there are more real negative samples for training and the effect of synthetic negative samples goes weak. (3) We find a counterintuitive result that the performance reaches best under the negative sampling ratio $0.3$ rather than $1.0$. We leave it for future research.

\subsection{Hyperparameter}
\begin{figure}
    \centering
    \includegraphics[width=1.0\linewidth]{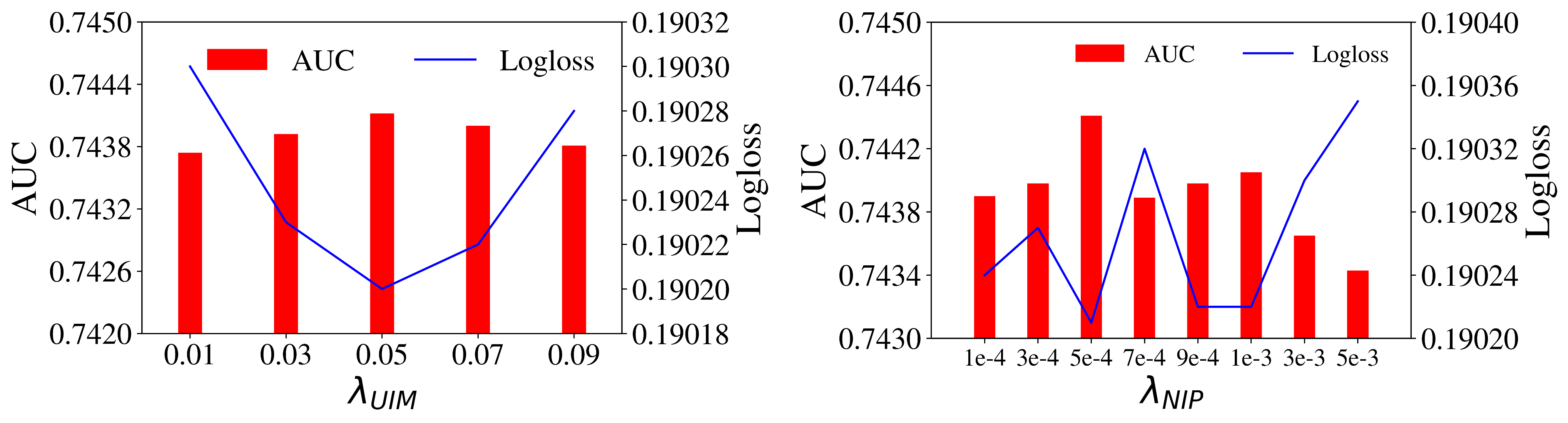}
    \caption{Result of various loss weights on Industry dataset.}
    \label{fig:ablation_industry_loss_weight}
\end{figure}

\begin{figure}
    \centering
    \includegraphics[width=1.0\linewidth]{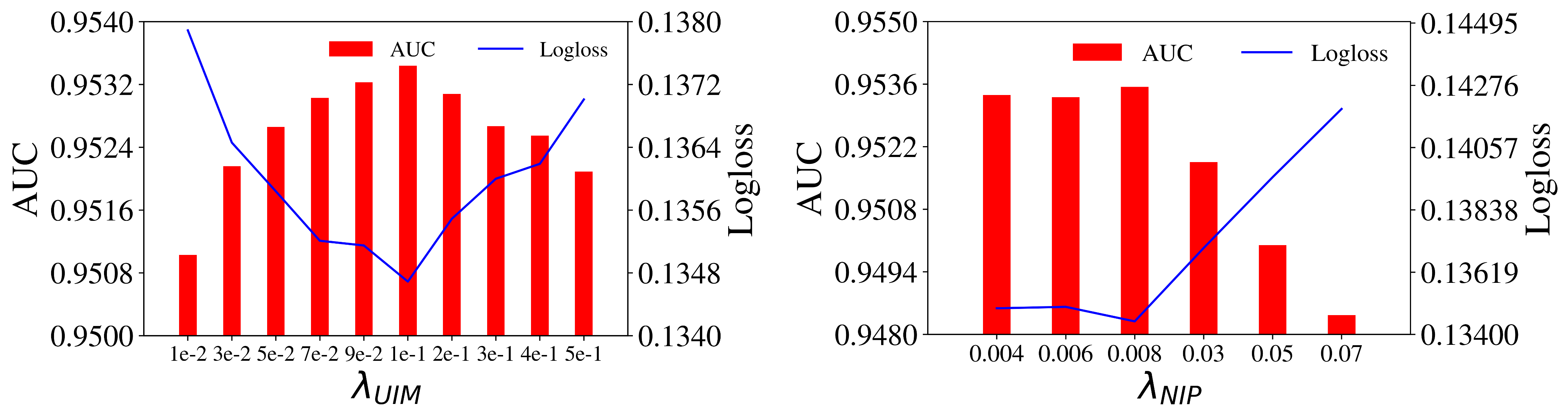}
    \caption{Result of various loss weights on Taobao dataset.}
    \label{fig:ablation_taobao_loss_weight}
\end{figure}

For hyperparameters, we study the effect of two auxiliary tasks' loss weight on both datasets. Figure~\ref{fig:ablation_industry_loss_weight} and Figure~\ref{fig:ablation_taobao_loss_weight} show the results of the two datasets respectively. For the loss weight of the UIM task, the improvements increase when the loss weight enlarges at the beginning, but then decrease when it further enlarges. The NIP is sensitive to the loss weight hyperparameter on both datasets. It needs to maintain the loss weight at a magnitude of small order otherwise the performance will fade away or even decline. Overall, properly tuned hyperparameters of loss weights can provide useful signals to accelerate the model's convergence. 

\subsection{Case Study}
\begin{figure}
    \centering
    \includegraphics[width=0.9\linewidth]{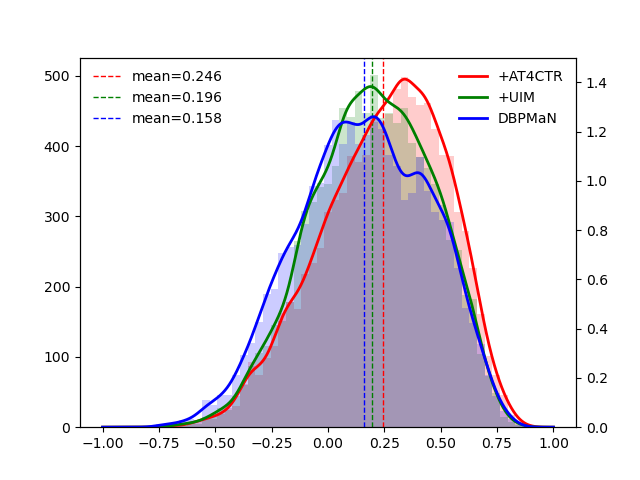}
    \caption{Relevance between user and item.}
    \label{fig:cosine}
\end{figure}
In this section, we study whether AT4CTR can strengthen the relevance between user and item of positive samples. We randomly select $10,000$ positive samples from the test set on the Taobao dataset. Then we extract the embedding of user side features and embeddings of item side features respectively. After that, we compute the cosine similarity of the concatenated embedding between user and item. We choose DBPMaN, DBPMaN$_{UIM}$, DBPMaN$_{AT4CTR}$ for analysis. The results in Figure~\ref{fig:cosine} demonstrate the two auxiliary match tasks can enhance the relevance between use and item. This reveals the deficiency of the main CTR task in learning the user-item relevance due to the data sparsity. AT4CTR makes up the deficiency and enhances the performance.

\subsection{Resource Cost}
In this section, we collect the statistical data to analyze the storage cost and the model training time on the industry dataset. From Table~\ref{tab:resource_cost}, the storage cost and the model training time increase rapidly as the negative sampling ratio increases. A large negative sampling ratio will influence the daily updating of the CTR model as the CTR model is only one component of the industry search advertisement system. Thus, negative down-sampling sometimes is unavoidable. But, AT4CTR just brings a little extra training time and no storage cost. When we take the negative sampling ratio as $0.1$, AT4CTR can obtain the same performance as the situation without negative down-sampling but save $6$ times storage cost and model training time. For the platform, saving costs can also increase revenue.

\begin{table}[tb!]
\centering
\caption{Resource cost of different negative sampling ratios. T means terabyte and h means hour.}
\label{tab:resource_cost}
\begin{tabular}{cl|c|c|c|c|c}
\toprule
& Ratio & \multicolumn{1}{c}{0.1} &\multicolumn{1}{c}{0.3} &\multicolumn{1}{c}{0.5} &\multicolumn{1}{c}{0.7}  &\multicolumn{1}{c}{1.0} \\
\midrule
\multicolumn{2}{c|}{Storage(T)}            & \multicolumn{1}{c}{39.0}   & \multicolumn{1}{c}{88.3}   & \multicolumn{1}{c}{137.0}  & \multicolumn{1}{c}{181.6}  & \multicolumn{1}{c}{230.2} \\
\midrule
\multicolumn{2}{c|}{Time(h)}               & \multicolumn{1}{c}{6.1}    & \multicolumn{1}{c}{13.9}   & \multicolumn{1}{c}{21.3}   & \multicolumn{1}{c}{29.1}        & \multicolumn{1}{c}{39.5} \\
\multicolumn{2}{c|}{Time$_{AT}$(h)}        & \multicolumn{1}{c}{6.4}    & \multicolumn{1}{c}{14.6}   & \multicolumn{1}{c}{22.5}       & \multicolumn{1}{c}{31.0}        & \multicolumn{1}{c}{-} \\
\bottomrule 
\end{tabular}
\end{table}

\subsection{Online Results}
We conduct an A/B test in the industry online search advertising system to measure the benefits of AT4CTR compared with the online baseline DBPMaN. The AT4CTR is allocated with 10\% serving traffic for one month. It achieves 1.27\% relative promotion on the Revenue Per Search and 7.21\% relative increase in the Return on Investment. 

\section{Conclusion}
In this paper, we propose the AT4CTR for enhancing the CTR model's performance. AT4CTR which contains UIM and NIP auxiliary tasks aims at alleviating the data sparsity problem by providing extra training signals. We conduct offline/online experiments to verify the
effectiveness of the AT4CTR. Finally, we do some ablation studies and visualization to show the correctness of AT4CTR’s component.

\section{Acknowledgemnts}
The work was supported by Meituan. Defu Lian was supported by grants from the National Natural Science Foundation of China (No. 62022077 and 61976198).

\bibliography{aaai24}

\begin{thebibliography}{32}
\providecommand{\natexlab}[1]{#1}

\bibitem[{Bian et~al.(2022)Bian, Wu, Ren, Pi, Zhang, Xiao, Sheng, Zhu, Chan, Mou et~al.}]{bian2022can}
Bian, W.; Wu, K.; Ren, L.; Pi, Q.; Zhang, Y.; Xiao, C.; Sheng, X.-R.; Zhu, Y.-N.; Chan, Z.; Mou, N.; et~al. 2022.
\newblock CAN: feature co-action network for click-through rate prediction.
\newblock In \emph{Proceedings of the fifteenth ACM international conference on web search and data mining}, 57--65.

\bibitem[{Chang et~al.(2023)Chang, Zhang, Fu, Zang, Guan, Lu, Hui, Leng, Niu, Song et~al.}]{chang2023twin}
Chang, J.; Zhang, C.; Fu, Z.; Zang, X.; Guan, L.; Lu, J.; Hui, Y.; Leng, D.; Niu, Y.; Song, Y.; et~al. 2023.
\newblock TWIN: TWo-stage Interest Network for Lifelong User Behavior Modeling in CTR Prediction at Kuaishou.
\newblock \emph{arXiv preprint arXiv:2302.02352}.

\bibitem[{Chen et~al.(2020)Chen, Kornblith, Norouzi, and Hinton}]{chen2020simple}
Chen, T.; Kornblith, S.; Norouzi, M.; and Hinton, G. 2020.
\newblock A simple framework for contrastive learning of visual representations.
\newblock In \emph{International conference on machine learning}, 1597--1607. PMLR.

\bibitem[{Cheng et~al.(2016)Cheng, Koc, Harmsen, Shaked, Chandra, Aradhye, Anderson, Corrado, Chai, Ispir et~al.}]{cheng2016wide}
Cheng, H.-T.; Koc, L.; Harmsen, J.; Shaked, T.; Chandra, T.; Aradhye, H.; Anderson, G.; Corrado, G.; Chai, W.; Ispir, M.; et~al. 2016.
\newblock Wide \& deep learning for recommender systems.
\newblock In \emph{Proceedings of the 1st workshop on deep learning for recommender systems}, 7--10.

\bibitem[{Dong et~al.(2023)Dong, Yu, Zhang, Lv, Wang, Jin, Wang, Wang, and Wang}]{dong2023deep}
Dong, J.; Yu, Y.; Zhang, Y.; Lv, Y.; Wang, S.; Jin, B.; Wang, Y.; Wang, X.; and Wang, D. 2023.
\newblock A Deep Behavior Path Matching Network for Click-Through Rate Prediction.
\newblock \emph{arXiv preprint arXiv:2302.00302}.

\bibitem[{Gao, Yao, and Chen(2021)}]{gao2021simcse}
Gao, T.; Yao, X.; and Chen, D. 2021.
\newblock Simcse: Simple contrastive learning of sentence embeddings.
\newblock \emph{arXiv preprint arXiv:2104.08821}.

\bibitem[{Guo et~al.(2017)Guo, Tang, Ye, Li, and He}]{guo2017deepfm}
Guo, H.; Tang, R.; Ye, Y.; Li, Z.; and He, X. 2017.
\newblock DeepFM: a factorization-machine based neural network for CTR prediction.
\newblock \emph{arXiv preprint arXiv:1703.04247}.

\bibitem[{Guo et~al.(2019)Guo, Hua, Jia, Zhao, Wang, and Cui}]{guo2019buying}
Guo, L.; Hua, L.; Jia, R.; Zhao, B.; Wang, X.; and Cui, B. 2019.
\newblock Buying or browsing?: Predicting real-time purchasing intent using attention-based deep network with multiple behavior.
\newblock In \emph{Proceedings of the 25th ACM SIGKDD international conference on knowledge discovery \& data mining}, 1984--1992.

\bibitem[{Guo et~al.(2022)Guo, Zhang, He, Qin, Guo, Chen, Tang, He, and Zhang}]{guo2022miss}
Guo, W.; Zhang, C.; He, Z.; Qin, J.; Guo, H.; Chen, B.; Tang, R.; He, X.; and Zhang, R. 2022.
\newblock Miss: Multi-interest self-supervised learning framework for click-through rate prediction.
\newblock In \emph{2022 IEEE 38th international conference on data engineering (ICDE)}, 727--740. IEEE.

\bibitem[{He et~al.(2014)He, Pan, Jin, Xu, Liu, Xu, Shi, Atallah, Herbrich, Bowers et~al.}]{he2014practical}
He, X.; Pan, J.; Jin, O.; Xu, T.; Liu, B.; Xu, T.; Shi, Y.; Atallah, A.; Herbrich, R.; Bowers, S.; et~al. 2014.
\newblock Practical lessons from predicting clicks on ads at facebook.
\newblock In \emph{Proceedings of the eighth international workshop on data mining for online advertising}, 1--9.

\bibitem[{Lian et~al.(2018)Lian, Zhou, Zhang, Chen, Xie, and Sun}]{lian2018xdeepfm}
Lian, J.; Zhou, X.; Zhang, F.; Chen, Z.; Xie, X.; and Sun, G. 2018.
\newblock xdeepfm: Combining explicit and implicit feature interactions for recommender systems.
\newblock In \emph{Proceedings of the 24th ACM SIGKDD international conference on knowledge discovery \& data mining}, 1754--1763.

\bibitem[{Lin et~al.(2023)Lin, Qu, Guo, Dai, Tang, Yu, and Zhang}]{lin2023map}
Lin, J.; Qu, Y.; Guo, W.; Dai, X.; Tang, R.; Yu, Y.; and Zhang, W. 2023.
\newblock MAP: A Model-agnostic Pretraining Framework for Click-through Rate Prediction.
\newblock In \emph{Proceedings of the 29th ACM SIGKDD Conference on Knowledge Discovery and Data Mining}, 1384--1395.

\bibitem[{Lin et~al.(2022)Lin, Zhou, Wang, Da, Chen, and Wang}]{lin2022sparse}
Lin, Q.; Zhou, W.-J.; Wang, Y.; Da, Q.; Chen, Q.-G.; and Wang, B. 2022.
\newblock Sparse Attentive Memory Network for Click-through Rate Prediction with Long Sequences.
\newblock In \emph{Proceedings of the 31st ACM International Conference on Information \& Knowledge Management}, 3312--3321.

\bibitem[{Lyu et~al.(2020)Lyu, Dong, Huo, and Ren}]{lyu2020deep}
Lyu, Z.; Dong, Y.; Huo, C.; and Ren, W. 2020.
\newblock Deep match to rank model for personalized click-through rate prediction.
\newblock In \emph{Proceedings of the AAAI Conference on Artificial Intelligence}, volume~34, 156--163.

\bibitem[{Mikolov et~al.(2013)Mikolov, Sutskever, Chen, Corrado, and Dean}]{mikolov2013distributed}
Mikolov, T.; Sutskever, I.; Chen, K.; Corrado, G.~S.; and Dean, J. 2013.
\newblock Distributed representations of words and phrases and their compositionality.
\newblock \emph{Advances in neural information processing systems}, 26.

\bibitem[{Oord, Li, and Vinyals(2018)}]{oord2018representation}
Oord, A. v.~d.; Li, Y.; and Vinyals, O. 2018.
\newblock Representation learning with contrastive predictive coding.
\newblock \emph{arXiv preprint arXiv:1807.03748}.

\bibitem[{Ouyang et~al.(2019)Ouyang, Zhang, Ren, Qi, Liu, and Du}]{ouyang2019representation}
Ouyang, W.; Zhang, X.; Ren, S.; Qi, C.; Liu, Z.; and Du, Y. 2019.
\newblock Representation learning-assisted click-through rate prediction.
\newblock \emph{arXiv preprint arXiv:1906.04365}.

\bibitem[{Pi et~al.(2020)Pi, Zhou, Zhang, Wang, Ren, Fan, Zhu, and Gai}]{pi2020search}
Pi, Q.; Zhou, G.; Zhang, Y.; Wang, Z.; Ren, L.; Fan, Y.; Zhu, X.; and Gai, K. 2020.
\newblock Search-based user interest modeling with lifelong sequential behavior data for click-through rate prediction.
\newblock In \emph{Proceedings of the 29th ACM International Conference on Information \& Knowledge Management}, 2685--2692.

\bibitem[{Rendle(2010)}]{rendle2010factorization}
Rendle, S. 2010.
\newblock Factorization machines.
\newblock In \emph{2010 IEEE International conference on data mining}, 995--1000. IEEE.

\bibitem[{Richardson, Dominowska, and Ragno(2007)}]{richardson2007predicting}
Richardson, M.; Dominowska, E.; and Ragno, R. 2007.
\newblock Predicting clicks: estimating the click-through rate for new ads.
\newblock In \emph{Proceedings of the 16th international conference on World Wide Web}, 521--530.

\bibitem[{Vaswani et~al.(2017)Vaswani, Shazeer, Parmar, Uszkoreit, Jones, Gomez, Kaiser, and Polosukhin}]{vaswani2017attention}
Vaswani, A.; Shazeer, N.; Parmar, N.; Uszkoreit, J.; Jones, L.; Gomez, A.~N.; Kaiser, {\L}.; and Polosukhin, I. 2017.
\newblock Attention is all you need.
\newblock \emph{Advances in neural information processing systems}, 30.

\bibitem[{Wang and Liu(2021)}]{wang2021understanding}
Wang, F.; and Liu, H. 2021.
\newblock Understanding the behaviour of contrastive loss.
\newblock In \emph{Proceedings of the IEEE/CVF conference on computer vision and pattern recognition}, 2495--2504.

\bibitem[{Wang et~al.(2023)Wang, Wang, Li, Gu, Lu, Zhang, and Gu}]{wang2023cl4ctr}
Wang, F.; Wang, Y.; Li, D.; Gu, H.; Lu, T.; Zhang, P.; and Gu, N. 2023.
\newblock CL4CTR: A Contrastive Learning Framework for CTR Prediction.
\newblock In \emph{Proceedings of the Sixteenth ACM International Conference on Web Search and Data Mining}, 805--813.

\bibitem[{Wang et~al.(2017)Wang, Fu, Fu, and Wang}]{wang2017deep}
Wang, R.; Fu, B.; Fu, G.; and Wang, M. 2017.
\newblock Deep \& cross network for ad click predictions.
\newblock In \emph{Proceedings of the ADKDD'17}, 1--7.

\bibitem[{Wang et~al.(2021)Wang, Shivanna, Cheng, Jain, Lin, Hong, and Chi}]{wang2021dcn}
Wang, R.; Shivanna, R.; Cheng, D.; Jain, S.; Lin, D.; Hong, L.; and Chi, E. 2021.
\newblock Dcn v2: Improved deep \& cross network and practical lessons for web-scale learning to rank systems.
\newblock In \emph{Proceedings of the web conference 2021}, 1785--1797.

\bibitem[{Xie et~al.(2022)Xie, Sun, Liu, Wu, Gao, Zhang, Ding, and Cui}]{xie2022contrastive}
Xie, X.; Sun, F.; Liu, Z.; Wu, S.; Gao, J.; Zhang, J.; Ding, B.; and Cui, B. 2022.
\newblock Contrastive learning for sequential recommendation.
\newblock In \emph{2022 IEEE 38th international conference on data engineering (ICDE)}, 1259--1273. IEEE.

\bibitem[{Zhang et~al.(2023)Zhang, Liu, Jiang, Wang, Zhuang, Wu, Gao, and Chen}]{zhang2023fairlisa}
Zhang, Z.; Liu, Q.; Jiang, H.; Wang, F.; Zhuang, Y.; Wu, L.; Gao, W.; and Chen, E. 2023.
\newblock FairLISA: Fair User Modeling with Limited Sensitive Attributes Information.
\newblock In \emph{Thirty-seventh Conference on Neural Information Processing Systems}.

\bibitem[{Zhou et~al.(2018{\natexlab{a}})Zhou, Bai, Song, Liu, Zhao, Chen, and Gao}]{zhou2018atrank}
Zhou, C.; Bai, J.; Song, J.; Liu, X.; Zhao, Z.; Chen, X.; and Gao, J. 2018{\natexlab{a}}.
\newblock Atrank: An attention-based user behavior modeling framework for recommendation.
\newblock In \emph{Proceedings of the AAAI conference on artificial intelligence}, volume~32.

\bibitem[{Zhou et~al.(2021)Zhou, Ma, Zhang, Zhou, and Yang}]{zhou2021contrastive}
Zhou, C.; Ma, J.; Zhang, J.; Zhou, J.; and Yang, H. 2021.
\newblock Contrastive learning for debiased candidate generation in large-scale recommender systems.
\newblock In \emph{Proceedings of the 27th ACM SIGKDD Conference on Knowledge Discovery \& Data Mining}, 3985--3995.

\bibitem[{Zhou et~al.(2019)Zhou, Mou, Fan, Pi, Bian, Zhou, Zhu, and Gai}]{zhou2019deep}
Zhou, G.; Mou, N.; Fan, Y.; Pi, Q.; Bian, W.; Zhou, C.; Zhu, X.; and Gai, K. 2019.
\newblock Deep interest evolution network for click-through rate prediction.
\newblock In \emph{Proceedings of the AAAI conference on artificial intelligence}, volume~33, 5941--5948.

\bibitem[{Zhou et~al.(2018{\natexlab{b}})Zhou, Zhu, Song, Fan, Zhu, Ma, Yan, Jin, Li, and Gai}]{zhou2018deep}
Zhou, G.; Zhu, X.; Song, C.; Fan, Y.; Zhu, H.; Ma, X.; Yan, Y.; Jin, J.; Li, H.; and Gai, K. 2018{\natexlab{b}}.
\newblock Deep interest network for click-through rate prediction.
\newblock In \emph{Proceedings of the 24th ACM SIGKDD international conference on knowledge discovery \& data mining}, 1059--1068.

\bibitem[{Zhu et~al.(2018)Zhu, Li, Zhang, Li, He, Li, and Gai}]{zhu2018learning}
Zhu, H.; Li, X.; Zhang, P.; Li, G.; He, J.; Li, H.; and Gai, K. 2018.
\newblock Learning tree-based deep model for recommender systems.
\newblock In \emph{Proceedings of the 24th ACM SIGKDD International Conference on Knowledge Discovery \& Data Mining}, 1079--1088.

\end{thebibliography}

\end{document}